\documentstyle[preprint,aps]{revtex}
\begin{document}
\title{ Gamow-Teller strength distributions for nuclei in
pre-supernova stellar cores }
\author{F. K. Sutaria and A. Ray}
\address{ Theoretical Astrophysics Group, Tata Institute of Fundamental
         Research, Bombay-400 005, India }
\address{ E-mail: fks@tifrvax.tifr.res.in , akr@tifrvax.tifr.res.in}
\maketitle
\begin{abstract}
\noindent
\label{sec:abstract}
 Electron-capture and $\beta$-decay of nuclei in the core of massive
stars play an important role in the stages leading to a type II
supernova explosion. Nuclei in the f-p shell are particularly
important for these reactions in the post Silicon-burning stage of a
presupernova star. In this paper, we characterise the energy
distribution of the Gamow-Teller Giant Resonance (GTGR) for
mid-fp-shell nuclei in terms of a few shape parameters, using data
obtained from high energy, forward scattering (p,n) and (n,p)
reactions. The energy of the GTGR centroid $E_{GT}$ is further
generalised as function of nuclear properties like  mass number,
isospin and other shell model properties of the nucleus.  Since a
large fraction of the GT strength lies in the GTGR region, and the
GTGR is accessible for weak transitions taking place at energies
relevant to the cores of presupernova and collapsing stars, our
results are relevant to the study of important $e^-$-capture and
$\beta$-decay rates of arbitrary, neutron-rich, f-p shell nuclei in
stellar cores. Using the observed GTGR and Isobaric Analog States
(IAS) energy  systematics we compare the coupling coefficients in the
Bohr-Mottelson two particle interaction Hamiltonian for different
regions of the Isotope Table.
\noindent
\vskip 0.01 true in
Ms number ~~~~~~~~~PACS number: 97.10.Cv, 23.40.-S, 97.60.Bw\\
\end{abstract}
\vfill
\eject
\section {Introduction}
\label{sec: Intro}
\noindent

It is well known \cite{BBB},\cite{BBAL} that during stellar collapse,
the final mass of the homologously collapsing core and the strength
of the subsequent type II supernova shock is determined by the final
electron fraction of the core ${Y_{ef}}$. The latter, in turn, is
influenced by the electron captures (and $\beta$-decays) taking place
on the nuclei present in the core during it's hydrostatic evolution,
Kelvin-Helmholtz contraction and dynamic collapse phases. This can be
seen from the scaling relations \cite{BBB},\cite{BBAL} for the mass
of the homologously collapsing core and for the shock energy when the
core bounces at supernuclear densities:-
$$ M_{HC} \propto Y_{ef}^2 $$
and
$$ E_{shock}\simeq {GM_{HC}^2 \over R_{HC}} (Y_{ef} - Y_{ei})
\simeq M_{HC}^{5/3} (Y_{ef} - Y_{ei})
\simeq Y_{ef}^{10/3} (Y_{ef} - Y_{ei}) $$
 where $M_{HC}$ and $R_{HC}$ are respectively the mass and radius of
the homologously collapsing core.  The electron-captures and
$\beta$-decays also determine the physical conditions in the (quasi-)
hydrostatically evolving core through their influence on the entropy
per nucleon $S_i$. This is because a low value of $S_i$ implies that
(a) the number of free protons is smaller and (b) the nuclei are
excited generally to low energy states. Since the core loses energy
(through neutrino emission) mainly by electron captures on free
protons, as a consequence of (a), the electron fraction $Y_e$ remains
high leading to a more massive homologously collapsing core and
subsequently a more energetic shock.  Further, since the collapse is
essentially adiabatic and the nuclei have low excitation energies,
the low value of $S_i$ ensures that the number of drip nucleons is
very small, i.e. the nucleons remain inside the nuclei right until
the core reaches nuclear density.  Thus since the number of free
nucleons is low, the collapse proceeds to higher densities leading to
a stronger bounce shock.  For all these reasons, to make a realistic
model of the physical conditions prevailing in the core of a type II
supernova progenitor, it is necessary to know the weak interaction
rates of a number of nuclei at the relevant energies.

  In general, the energy dependence of weak interaction matrix
elements (or equivalently, the Gamow-Teller (G-T) strength
distribution) is unknown for many nuclei of potential importance in
pre-supernova stars and collapsing cores. However for the few nuclei
of interest for which the experimental data is available, it is
possible to map out the energy distribution of the GT strength in
terms of a few shape parameters. These parameters can be related to
nuclear properties like the nuclear isospin and the mass number and a
corresponding approximate distribution can be constructed for any
nucleus of astrophysical interest in that nuclear shell.  We consider
here nuclei in the f-p shell since in the post-Silicon-burning phase
the core of a pre-supernova massive star has a significant abundance
of these neutron rich nuclei with ${A \ge 60}$. In e$^-$-capture and
$\beta$-decay calculations relevant to the pre-supernova scenario, it
is not particularly important to consider the higher shells (though
this is not the case during the collapse phase, where at sufficiently
high density they may have non-negligible abundances). This is
because neutron shell blocking \cite{Full82} (which starts at around
$^{74}$Ge) substantially decreases  the e$^-$-capture rate for these
heavier nuclei and their contribution to the collective electron
capture rates is small compared to that of mid f-p shell nuclei, if
the latter is present with sufficient abundances.

The importance of knowing the centroid of the GT distribution
($E_{GT}$) lies with the fact that it determines the effective energy
of the e$^-$-capture and $\beta$-decay reactions from e.g. the ground
state of the initial nucleus to the excited state of the final
nucleus, and this  along with the electron- Fermi energy  determines
which nuclei are able to capture electrons from, or $\beta$-decay
onto the Fermi-sea at a given temperature and density --- thus
controlling the rate at which the abundance of a particular nuclear
species would change in the pre-supernova core. For example, the
following expression holds for the $\beta^-$-decay rates of the
discrete states of the mother nucleus (see e.g. \cite{Karetal}):
\begin{eqnarray}
\lambda_s &= &{\rm ln 2} {(6250 sec)^{-1} \over G} \sum_i(2J_i+1)\exp
\left(-E_i/kT\right)  \nonumber \\
          &\times& \int^{Q_i}_0
          \left[ |M^i_F(E^{\prime})|^2 +
\left( {g_A \over g_V}\right)^2|M^i_{GT}(E^{\prime})|^2 \right]
f(Q_i-E^{\prime})dE^{\prime}
\end{eqnarray}
where
$$G=\sum_i (2J_i+1) \exp(-E_i/kT) $$
 $g_A$ and $g_V$ are the axial-vector and vector coupling constants
respectively and the $|M_{F,GT}^i|^2$ are the corresponding Fermi and
GT matrix elements. $E_i$ is the energy of the $i^{th}$ level of the
mother having spin $J_i$, and Q-value $Q_i=Q+E_i$. The function
$f(Q_i - E^{'})$ is the  phase-space factor, integrated over
electron-energies (in units of $ m_e$) from 1 to  $(Q_i - E^{'})$.
Eq. (1) shows that the distributions of the transition strengths in
energy are important in the rate calculations.

The $\beta$-decays can take place either through the Fermi (vector)
type interaction or the Gamow-Teller (axial-vector) type interaction
while the e$^-$-captures on a nucleus in ground state involves only
the GT interaction.  The GT operator $ G_{A} \sum_{i} \sigma_{i}
{t_{i}^\pm}$ does not commute with the strong spin, isospin dependent
forces of the nuclear Hamiltonian, which causes a mixing of states in
both the spin as well as isospin space. While the effect of mixing in
the isospin space is small (because of the relatively weaker Coulomb
potential term), the mixing of states in the spin-space gives rise to
a broad distribution of the total GT strength in excitation energies
known as the Gamow-Teller Giant Resonance (GTGR) which contains
a large fraction of the total weak-interaction strength sum. In
contrast, the Fermi strength is concentrated in a narrow region of
excitation energies because the Fermi operator commutes with all
parts of the nuclear Hamiltonian except the Coulomb part. At
temperatures and  densities (${T \simeq 0.5 MeV}$ and ${\rho \simeq
10^{10} g/{cm^3}}$ ) characteristic of the presupernova core, the
main contribution to the weak interaction rates is expected to come
from the GTGR region.

In sections~\ref{sec: GTstrength}  and~\ref {sec:GTcentroid} we use
the available data on nuclear charge-exchange reactions on mid
fp-shell nuclei to characterise the GT strength distribution for
arbitrary nuclei of astrophysical interest. In section~\ref{sec:
Comp} we compare our results with other theoretical methods, namely
the M1 method used by Klapdor \cite{Klap76} and the method of Fuller,
Fowler and Newman \cite{FFNII}.  In section~\ref{sec: CConst} we
discuss the implications of the observed GT energy systematics
vis-\`a-vis the Bohr-Mottelson two body Hamiltonian and in
section~\ref{sec: Concl} we give our conclusions.

\section {The Gamow-Teller Strength Distribution}
\label{sec: GTstrength}
\noindent
It is well known \cite{Horetal},\cite{Gaaretal} that the information on
charge exchange reactions in nuclei can be used to extract the GT
strength distributions with respect to the excitation energy of the
daughter nucleus.  The charge-exchange (p,n) and (n,p) reactions are
iso-spin and spin-dependent over a wide range of projectile energies.
The weak interaction strength distribution ($B_{GT}^+$ or $B_{GT}^-$)
is propotional to the ${\Delta L = 0}$, forward scattering
cross-section in the high projectile energy, low momentum transfer
limit.  This can be seen by comparing the corresponding operators
\cite{Tadetal}:-
 $${  \sum_{i} V_{\sigma \tau}(r_{ip}) \sigma_i.\sigma_p \tau_i.
       \tau_p}\;
      {\rm and}\; { \sum_{i} V_{\tau} (r_{ip}) \tau_{i}.\tau_{p} }$$
 which are similar to:
 $${      G_{A} \sum_{i} \sigma_{i} {t_{i}^\pm}\;
 {\rm and}\;   G_V \sum_{i} t_i^\pm } \;.  $$
The ${\Delta L = 0}$  scattering cross-section for various mid-fp shell
target nuclei undergoing (p,n) and (n,p) reactions are reported in
ref. \cite{Rap83} to \cite{Alf90}. These cross-sections are extracted
out of the  $0 \deg$ spectrum  by means of multipole analysis using
Distorted Wave Impulse Approximation calculations (see
ref.\cite{Tadetal} and references therein).  The cross-sections can
be analysed to obtain the experimental distribution of GT strength in
energy to a reasonably good accuracy  by using known calibration
relations (e.g. the relation developed in \cite{Tad81} ) and this
method has been widely used in the literature
\cite{Rap83},\cite{Elkat}.

The calculation of the weak-interaction mediated reactions under the
astrophysically relevant conditions requires the knowledge of not
only the total strength in a given direction ( e.g. $\beta^-$-decay
or e$^-$-capture) but also that of the strength distribution in
nuclear excitation energy. Attempts have been made to obtain these
distributions theoretically from shell-model calculations. However,
the number of shell-model basis states can get very large for mid-fp
shell nuclei, even at low energies. Hence the direct method of
obtaining GTGR distribution from shell model calculation using full
$0 \hbar \omega$ basis is computationally very involved even for a
few nuclei.  Several attempts to calculate the GT distribution from
direct shell model calculations using a truncated shell model basis
space have also been made, (e.g. Aufderheide et al. \cite{Aufetal} )
which because of the above reasons, despite substantial computational
efforts, are as yet approximate. We note that complete fp-shell
$0\hbar  \omega$ {\it Monte Carlo} calculations have been mentioned
recently  which show significant quenching (discussed later), and
have reportedly reproduced  the experimentally observed strength in
nuclei such as $^{54}$Cr, $^{54}$Fe, $^{55}$Mn and $^{56}$Fe  but not
in $^{58}$Ni, the result being sensitive to the interaction
Hamiltonian assumed (see Koonin and Langanke \cite{Koonetal} and
references therein). In any case the calculation of the weak
interaction mediated reaction rates for a moderately large number of nuclei
required in astrophysical situations requires a straightforward and
computationally manageable approach.  Here, following an earlier work
\cite{Karetal} we use the framework of a statistical approach, as in
the Spectral Distribution Theory \cite{Frenetal},\cite{Kotaetal}.

The sum rule strength for a transition from an initial state
$|i\rangle$ to a final state $|f\rangle$ is given by
$$ S_{\bf \sigma \tau} = \sum_{f} |\langle f |{\bf \sigma \tau}|
     i \rangle |^2
   = \sum_{f}\langle i |({\bf \sigma \tau})^{\dagger}| f \rangle
       \langle f |{\bf \sigma \tau}| i \rangle
   =   \langle i |({\bf \sigma \tau})^{\dagger} {\bf \sigma \tau} |
     i \rangle
$$
On the single particle level, if $\langle n_{nlj^{'}}^P \rangle$ and
$\langle n_{nlj}^N \rangle $ are the fractional occupancies of the
neutrons in $nlj$ level and protons in ${nlj^{'}}$ level, then the GT
sum rule (e.g. for ${\beta}^-$-decay) is given by the expression:
$$  S_{\beta^-}^{GT} = 3 Z_n \sum_{nljj^{'}} | C_{nl}^{jj^{'}} |^2
                   ( 1 - \langle n_{nlj^{'}}^P \rangle)
                   \langle n_{nlj}^N \rangle  \eqno(2)$$
\noindent
where $C^{jj^{'}}_{nl} = (-1)^(j^{'} -j)[2(2j+1)(2j^{'}+1)]^{1/2}W(l
{1 \over 2} j 1; j^{'} {1 \over 2})$, and W is the Racah coefficient.
The occupation numbers $\langle n_{nlj^{'}}^P \rangle$, $\langle
n_{nlj}^N \rangle $ etc. in a given shell can be calculated with the
Spectral Distribution Theory ( see e.g. ref. \cite{Frenetal} and
\cite{Kotaetal})

For the distribution of the strength with the nuclear excitation
energy we note that the according to the Spectral Distribution
Theory, for a many particle space of large dimensionality, the
smoothed-out eigenvalue distribution of a (2+1) body Hamiltonian is
approximately a Gaussian. A skewed Gaussian (called the Edgeworth
expansion) of the form:
 $${ B_{GT}^{\pm} = A_0 [ 1 + \gamma_1 (x^3 - 3x)/6 + \gamma_1^2
                    (x^6 -15 x^4 + 45 x^2 -15)/ 72 ] \exp(-x^2 / 2) }$$

 where
 $${ x = (E_{ex} - E_{GT})/ \sigma}$$
is used here. The parameters $\gamma_1$ (the skewness factor),
$E_{GT}$ (the energy centroid), $\sigma$ (the effective half-width)
and $A_0$ (normalisation factor) are obtained for each daughter
nucleus by fitting the above formula to the experimentally obtained
strength distribution.

It is known that if the total strength in the $GT^+$ direction is
small, the observed value of the total Gamow-Teller strength in the
$GT^-$ direction  is, on the average, approximately ${50 \%}$ of the
theoretically predicted value of $3(N-Z)$ ( for the nuclei studied
here, it is observed to vary between 45 to ${63 \%}$). This quenching
of GT strength given by the factor $Z_n$ in eq. (2) above is expected
to be due to the excitation of $\Delta$-isobars at higher
excitation energies through $N^{-1} - \Delta $ transitions, or the
missing GT strength may lie at higher excitations (between 30 MeV to
50 MeV) \cite{Gaaretal}.  On the other hand in the electron capture
direction, the total GT strength sum ( i.e.
$\int_0^{\infty}{B_{GT}^+}(E^{'}) dE^{'}$ ) in the f-p shell has been
argued to depend on the number of  valance protons and the number of
neutron holes in the fp shell (Koonin and Langanke, 1994). According
to these authors the {\it total} ${GT}^+$ strength observed
experimentally in the mid-fp shell nuclei can be fitted by an
expression like:
$$ B(GT_+)= 0.0429 Z_{val}( 20 - N_{val}). $$
As argued by them, in the electron capture direction for nuclei
having $N>Z$, the number of possible transitions are limited by the
number of available neutron holes and the number of protons in the
full f-p shell. They also suggest that the neutron-proton
correlations throughout all of the f-p shell are important in
reproducing the observed quenching, and mid-fp shell nuclei in this
sense behave as though there is only one large shell in which all
subshell structure has been diluted. The above expression includes
the observed quenching and can therefore be used to independently fix
$Z_n$ for the sum-rule for electron capture in an equation similar to
eq. (2)

For reasons discussed in section~\ref{sec: Intro} and the beginning
of this section, astrophysical calculations require, apart from the
total $GT^+$ strength itself (such as given above), it's distribution
in energy --- which is characterised by higher moments of the
distribution such as the energy centroid E$_{GT}$, the skewness of
the distribution, etc.  With the objective of generating the
$B_{GT}(E)$ distribution of any arbitrary nucleus (A,Z) of potential
astrophysical importance, we now make use of the experimentally
available data. The $B_{GT}^+(E)$ and $B_{GT}^-(E)$ distributions
obtained from analysis of charge exchange experiments are
least-square fitted to the Edgeworth expression discussed above,
under the constraint that the total `experimental' GT strength is
reproduced by the area under the fitted expression. The Edgeworth
expansion gives a good fit in the cases of $^{51}$V, $^{81}$Br,
$^{71}$Ga and $^{60}$Ni.  In the case of $^{58}$Ni (target nucleus),
the resonance region between 7.5 and 12 MeV is fitted well by the
Edgeworth expression.  Similarly for $^{54}$Fe, the fit is reasonably
good between 6 to 12 MeV. For these nuclei, the strength fluctuates
rapidly at low excitation energy, suggesting that at these energies
single particle transitions dominate over collective resonance. Since
Spectral Distribution Theory is based on a statistical description of
a large number of basis states, it is not expected that the Edgeworth
expansion will reproduce the fluctuating strengths at low energies
due to single particle transitions. Nevertheless for the G-T
collective resonance region (which is important because of the bulk
of the total strength being accessible at energies relevant to the
astrophysical situation), the Spectral Distribution Theory  approach
is useful.

Finally, since for $^{56}$Fe, $^{54}$Fe, $^{58}$Ni, $^{60}$Ni, only
the $\sigma(L=0)$ (p,n) cross-section is reported in ref
\cite{Rap83}, the calibration relation based on the factorised DWIA
expression for $L=0$ differential cross-section as developed in
\cite{Tad81} was used to convert the reported cross-section to
${B_{GT}^-}/MeV$. Table 1 gives the Edgeworth parameters for each of
these fits.  A typical fit of $B_{GT}^-$ per $0.1$ MeV interval (for
$^{71}$Ga) is shown in Fig. 1. The data used in these fits came from
refs. \cite{Rap83} ($^{54,56}$Fe, $^{58,60}$Ni), \cite{Krof87}
($^{81}$Br), \cite{Rap84} ($^{51}$V) and \cite{Krof85} ($^{71}$Ga).

The Edgeworth parameters for $B_{GT}^+(E)$ per MeV  vs. $E_x$ fits are
tabulated in Table~\ref{tab: Edge+}. Fig. 2 shows the distribution
and fit for $^{59}$Co. The distribution is well reproduced for the
low energy region and the main GTGR peak.  For this set of results
the data was obtained from refs. \cite{Vet89} ($^{54}$Fe),
\cite{Elkat} ($^{56}$Fe, $^{58}$Ni, $^{55}$Mn), \cite{Alf93}
($^{51}$V, $^{59}$Co), \cite{Vet92} ($^{70}$Ge).
For $^{70}$Ge, Vetterli et al. \cite{Vet92} quote the value of
$B_{GT}^+$ upto 7.6 MeV only, because of the difficulty
of extraction of the $\Delta L=0$ component from the (p,n) cross-
section. The parameters for this nucleus reported in
Table~\ref{tab: Edge+} are obtained on the basis of the strength
upto 7.6 MeV only.
In some cases the single-particle
transitions  may dominate in either direction at low excitation
energy, and so wherever they are known experimentally, these should
be used explicitly in the transition rate  calculations.

\section{Predicted centroids of GT strength distributions}
\label{sec:GTcentroid}
\noindent
We refer to the centroids of the $B_{GT}^\pm(E)$ distributions
obtained by making Edgeworth fits to those reported from the (p,n)
and (n,p) experiments as the `experimental centroids'. In order to be
able to predict such an energy distribution for any relevant f-p
shell nucleus, we relate the shape-parameters  for $B_{GT}^\pm$
distributions to nuclear properties like the nuclear ground-state
isospin, and the mass number. Here we report the relation developed
for the energy centroid $E_{GT}$, which can be applied to arbitrary
nuclei of astrophysical interest for rate calculations.
\subsection{ $GT^- $ energy systematics}
\noindent

 The GT$^\pm$ operator is a space vector and an isovector, and the
selection governing this transition require that ${\Delta J= 0 , \pm
1}$, ${\Delta \pi=0 }$ (no $0^+ \rightarrow 0^+$ )
and ${\Delta T= 0, \pm 1}$. Since the Fermi operator is a pure
isovector, the selection rules require that ${\Delta J = 0}$,
${\Delta \pi =0 }$ and ${\Delta T= 0}$, i.e.
transitions take place only between Isobaric Analog States (IAS).
For both Fermi and allowed GT transitions ${\Delta L =0}$. It
has been argued (see \cite{Naketal}, \cite{Horetal} and \cite{FFNII})
that under the action of resonant GT or Fermi operators, the
collective states have centroids located at $E_{GT}^-$ and $E_{IAS}$
with respect to the appropriate, unperturbed state such that the
difference $E_{GT^-}-E_{IAS}$ should depend on the spin-orbit
splitting term  ($\sim A^{-1/3}$), and on the isospin dependent Lane
potential term ($\sim (N-Z)/A $). The latter gives the energy
difference between two levels having the same $T_z$, but differ in
$T$ by a one unit. The excitation energy of the Isobaric Analog State
$E_{IAS}$ (where the Fermi centroid is located) is obtained in our
analysis, either experimentally (where available), or from the
following theoretical relation developed by Fowler and Woosley and
reported in ref \cite{FFNII}  for neutron rich nuclei :
  $${ E_{IAS} = \Delta M_A - \Delta M_C - 0.7824 + {1.728 (Z_C -1)\over R}
       MeV}  \eqno(3)$$
with
  $${ R=1.12 A^{1/3}+0.78} $$
 Here $\Delta M_A$ is the mass excess of the initial nucleus, $\Delta
M_C$ is the mass excess of the final nucleus, and R is the radius in
fm. This is observed to agree very well with the experimental data
used in this work (see Fig. 3). Fitting a linear combination of the
spin-orbit and isospin dependent terms to the experimental centroids,
$E_{GT}^-$ for the (p,n) reaction data, we get the following
empirical relation :-
$${ E_{GT^-}-E_{IAS}= 44.16 A^{-1/3} - 76.1 (N-Z)/A } .\eqno(4)$$
The correlation between the theoretical $E_{{GT}^-}$ obtained from
(4) (together with the experimental values of E$_{IAS}$ reported in
the references given earlier), and the experimental GT$^{-}$
centroids as referred to above is shown in Fig. 4. It is noted that a
similar relation derived earlier \cite{Naketal} (also in
\cite{Horetal}) using data for nuclei well beyond f-p shell had the
corresponding coefficients differ substantially from those obtained
in eq. (4) for the f-p shell nuclei. The reason for this difference
is discussed in section~\ref{sec: CConst}. The goodness of the fit is
measured in terms of the rms deviation from the experimental value
and is evaluated to be 0.43 MeV. As it is apparent from
Table~\ref{tab: FFNcomp} and Fig. 4, the GT$^-$ centroids are located
roughly between 9 to 11 MeV. This, together with the expected thermal
spread of the electron Fermi distribution at temperatures relevant in
the presupernova core (typically $5 \times 10^9 K$), and a $ \sigma_N
\simeq 3$ MeV would make a typical error of the order of 0.43 MeV in
the predicted centroid  quite acceptable for astrophysical rate
calculations.
\subsection{ $GT^+$ energy systematics }
\noindent

Since in the n-p reaction there is a transition from a parent nuclear
ground-state having isospin $T=T_z$ to a daughter state having
minimum isospin $T_d=T_z + 1$ (for g.s. to g.s. transition ), there
can be no IAS state in the daughter corresponding to the ground state
of the mother for the electron capture or the $GT^+$ direction. Thus,
$E_{GT^+}$ has to be obtained directly (in contrast to
$E_{GT^-}-E_{IAS}$ in the $GT^-$ direction) and can be expected to
depend on the spin-orbit splitting term, the Lane potential term and,
for odd-A nuclei on the pairing energy term as well.

Electron-capture/e$^+$-emission on a nucleus generally raise the
final nucleus to an excited state. The single-particle excitation
configurations of most of these excited states can be constructed
from its ground state by breaking either a proton or a neutron pair
and raising a single particle to an excited level or by raising an
unpaired single particle (if available) to a excited state. For
odd-even/even-odd nuclei, the excited states of the final
even-odd/odd-even  nuclei (which are connected to ground state of the
parent nucleus by $GT^+$ transitions), can be generated from the g.s.
of the final nucleus by breaking a particle pair in lower energy
level and using it to create an unpaired particle in a higher energy
level or to pair off a previously unpaired particle in a higher
energy level and simultaneously raise another single particle to the
excited state.  For an odd-odd / even-even initial nucleus, however,
the excited states in the final even-even/ odd-odd nucleus will be
generated by breaking and making a particle pair. This  implies that
the energy required to break a pair in the final odd  A nucleus must
also be included in any expression for $GT^+$ energy systematics.

Fitting the experimental $E_{GT}^+$ centroids (obtained from the
Edgeworth weighted experimental $B_{GT}^+(E)$ distributions) to a
linear combination of these terms, we arrive at the following
empirical relation for $E_{{GT}^+}$ centroid with respect to the
ground state of the final nucleus:
 $${ E_{GT^+}= 13.10 A^{-1/3} - 11.28 (N-Z)/A + 12 A^{-1/2}
   \delta_{A_{odd}} } .\eqno(5)$$
 The plot of  $E_{GT^+}$ from (5) vs. the experimental centroid is
shown in Fig. 5. The goodness of fit in Fig. 5 is also measured in
terms of the r.m.s. deviation from the experimental value, and has
the value 0.31 MeV. The nuclei used in both figures are stated in the
order of increasing $E_{GT}$(experimental).  We note that Koonin and
Langanke (1994) suggest that the GT-resonance appears in (n,p)
spectra at systematically higher energies for odd-Z targets than for
even-Z  targets, although this is based on data involving target
nuclei with even number of neutrons. On the basis of single particle
excitation diagrams for the final nucleus, conforming to G-T
selection rules for a transition from the initial to the final
nucleus,  we find that the pairing energy dependence in terms of
odd-A in eq. (5)  (the last term) would  be present even in odd-N,
even-Z nuclei.

 The average difference of experimental and predicted values of 0.4
MeV in $E_{GT}^-$ and 0.3 MeV in the case of $E_{GT}^+$ can be
mitigated by the spread in the electron Fermi-Dirac distribution due
to the 0.5 MeV temperature and also due to the fact that in a given
region of the star at a particular stage of evolution there are
several nuclear species present in the stellar material
simultaneously. The total transition rate of this large admixture of
nuclei is likely  to be dominated in a given stage and region by a
few nuclear species, with relatively large transition rates and
abundances. These would necessarily have large energies of transition
and would in any case consume a substantial fraction of the strength
sum rule. Thus for astrophysical purposes, the level of accuracy of
the predictions for the GT$^\pm$ centroids obtained here are expected
to be adequate.

Data for $^{45}$Sc and $^{48}$Ti are also available in references
\cite{Alf91} and \cite{Alf90}. However, we have not used these
two nuclei in obtaining eq.(5) and Fig. 5, partly because these
nuclei are close to the beginning of the fp-shell where the general
assumptions of the statistical nature of the Spectral Distribution
Theory may not be valid.  For this reason, the calculation of the
Edgeworth weighted centroid of the experimental B$_{GT^+}(E)$
distribution gives inaccurate results because of the low energy
strength. Indeed, if these nuclei were to be included in Fig. 3, the
goodness of fit would have decreased to an r.m.s. value of 1.37 MeV.
Similarly, we had to ignore the data for $^{90}$Zr \cite{Ray90}
because the $g_{9/2}$ shell is occupied in this nucleus.
Effectively, it can be said that eq. (5) is valid for mid-fp-shell
nuclei with $ N>Z $.

\section{ Comparison with other theoretical methods }
\label{sec: Comp}
\noindent

While a detailed shell model calculation for these nuclei requires a
very large number of basis states and would be difficult to carry out
for each of these nuclei, it is conceivable that a reasonably good
estimate of the GT excitation energy centroids by using either the M1
excitation method (for the $GT^+$ excitation) or the FFN method (for
the $GT^-$ excitation). These methods are described in detail in ref.
\cite{FFNII}.  The predictions of these methods are compared with the
results obtained in this paper in the following paragraphs and in
Table~\ref{tab: FFNcomp} and Table~\ref{tab: M1comp}.

 In the FFN method, the GT transition in $GT^+$ ($GT^-$) direction,
which can be either the spin-flip (sf) or the no-spin-flip (nsf)
transition, is reproduced by taking into account all possible
excitation of a p(n) in the parent nucleus to a sf or nsf orbital
(for which $\Delta L =0 $) and then converting the p(n) to n(p) via
the $\tau^+$($\tau^-$) operator. The GT excitation energy for that
transition is given by the sum of the excitation energies with
respect to the ground state of the daughter nucleus together with the
Lane potential energy and if required, the pairing energy term. The
GT centroid energy with respect to the ground state of the daughter
nucleus is given by the weighted average of all these sf and nsf
transitions.  For example, in Fig. 6, the ground state of (g.s.)
$^{51}$V is connected to the g.s. of $^{51}$Cr by a nsf-$\tau^-$
transition, and to an excited state by a sf-$\tau^-$ transition. For
the former transition, the GT excitation energy (measured, as always,
with respect to the ground state of $^{51}$Cr) equals the Lane
potential ( $ 50(N-Z)/A $), which is evaluated as 4.90 MeV. The
square of the GT matrix element for this transition is 6.42. For the
latter transition, the GT excitation energy is the sum of
single-particle excitation energies (6.66 MeV), the Lane potential
(4.90 MeV) and pairing energy (1.68 MeV), while the square of the
transition matrix element is 13.71.  Here, the sf transition has the
maximum contribution to the total strength and this is observed to be
the norm in the case of other nuclei considered in ref.\cite{FFNII}
as well. Thus, from this single particle excitation method, the
$GT^-$ centroid can be expected to lie at 10.58 MeV. The experimental
centroid lies at 11.75 MeV. However, as this method does not take
into account the collective particle-hole excitations
\cite{Gaaretal} of the nucleus (where a good fraction of the total
strength can be expected to lie), it should not be expected to yield
the experimental $E_{GT}$ precisely. This is a purely theoretical
method, and the excitation energies are calculated from
single-particle shell model energy levels \cite{Hill}.
For the single particle levels of neutron we have used the sequenece and
values labelled as ``Seeger" in ref \cite{Hill}. Although for the protons,
the ``Seeger" sequence lists 2p$_{3/2}$ higher than 1f$_{5/2}$, since the
observed ground state spins of $^{63}$Cu, $^{65}$Cu etc (where there is one
extra proton beyond the 1f$_{7/2}$ level) contradict this sequencing (i.e.
the g.s. J$^{\pi}$ for these nuclei is (3/2)$^{-}$ and not (5/2)$^{-}$),
following an earlier work \cite{Karetal} we adopt the same values for the
proton single particle energy differences as in the case of neutrons in this
work unless specifically mentioned otherwise.

Now, it is seen from the selection rules that most of the GT Strength
for nuclei with $A\simeq 60$ lie in the spin-flip transitions.
Experimentally too, it is observed that the sf transition strength is
more concentrated than the nsf transition strength. For the sf
transitions, the action of the GT operator is equivalent to a s-f
transition followed by an isospin-flip (M1 transition from the mother
to the daughter). If we consider a $T^<$ to $T^>$ transition, to find
$E_{GT^+}$ in the daughter by the M1 method, we need to consider the
M1-$\tau^-$ transition from the daughter to the mother nucleus. The
M1 method for finding the $GT^+$-centroid in the neutron rich
daughter nucleus (see e.g. \cite{FFNII} ) is based on the observation
that most of the GT sf-configuration strength is concentrated near
the Anti-Isobaric Analog State (AIAS)  of the  M1 giant
resonance---the so called M1-AIAS state. Now, the predominant
configuration in the M1-AIAS state in the daughter is the spin-flip
configuration generated from a $T^<$ mother ground state  by
transforming a proton in the $j=l+1/2$ state into an empty neutron
in the $j=l-1/2$ level. The spin-flip configuration in the daughter
is therefore the starting point and it is this configuration whose
excitation with respect to the g.s. would yield the sought for
GT-centroid. The M1 method generates, in the $T^<$ mother, the $T^>$
analog of this sf- configuration by the application of the $T^+ =
\sum_{i=1}^A \tau_i^+$  (isospin-raising) operator . In general,
there may be more than one AIAS of the sf-configuration which
are orthogonal to each other. These AIAS states contain most of
the M1 excitation configuration strength and their energy
can be calculated from the single-particle shell model, taking
into account the particle-hole repulsion energy and the pairing
energy wherever relevant.  The $T^>$ Analog state is seperated
upwards from the AIAS states by the Lane potential. Once the energy
of this $T^>$ Analog state in the mother ( of the sf-configuration
in the daughter ) is known, subtraction of the energy of the Analog
state corresponding to the g.s. of the daughter (`the first analog')
would yield the approximate excitation energy of the sf-configuration
in the $T^>$  daughter with respect to its g.s., which is the required
$GT^+$  centroid in the daughter.  This assumes an argument similar
to the Brink hypothesis, as discussed later. The first Analog state
in the mother is often known experimentally, and where available,
this is an advantage in the M1-method.

 As an illustration of this method, we consider the case of GT$^+$
centroid in $^{54}$Mn (for the $^{54}$Fe(e$^-$, ${\nu}_e$)$^{54}$Mn
reaction). Fig. 7a shows the g.s. configuration of the mother
nucleus $^{54}$Fe. The spin-flip excitation of $^{54}$Mn are shown in
Figs. 7b  and alongside it are shown the excited states of $^{54}$Fe
which are obtained by operating ${\tau}^-$ on the spin-flip state of
$^{54}$Mn.  Thus the GT excitation corresponding to the sf excitation
in Fig. 7b (in $^{54}$Mn)  will be a superposition of two shown basis
configurations. Now, the AIAS corresponding to these two
configurations can be constructed and its position in energy is
determined as the weighted sum of the single particle excitation
energies and the Lane potential (3.70 MeV for this case). The
weighting factors obtained from M1 method for the $^{54}$Fe excited
state configurations are  $\protect\sqrt{3/4}$ and
$\protect\sqrt{1/4}$ respectively (see Fig. 7) and the excitation
energies for each configuration in Fig. 7b are  noted below them.
The position of the GT excitation  corresponding to the s-f
configuration of  $^{54}$Mn will then be given by the difference in
the energy of the obtained IAS state and the IAS state in $^{54}$Fe
corresponding to the g.s. of $^{54}$Mn (the so called first analog
state---located at an excitation energy of 8.7 MeV---see below).

A few examples of the M1 method in $T^<$ to $T^>$ transitions are
given in Table~\ref{tab: M1comp}. For all the nuclei quoted in
Table~\ref{tab: M1comp}, the required $E_{IAS}$ in the mother is
known from experiment. For the mother nuclei quoted in
Table~\ref{tab: FFNcomp} the $E_{IAS}$  of the daughter ground states
were obtained from \cite{NDS51} ($^{56}$Fe) and \cite{NDS69}
($^{59}$Co) and are 11.51 and 9.55 MeV respectively. For $^{54}$Fe we
note that ref. \cite{NDS50} gives only the Isobaric Analog States
corresponding to a few excited levels of $^{54}$Mn. The difference of
the IAS energies and the corresponding levels in $^{54}$Mn are in the
range of 8.59 to 8.79 MeV.  Using the assumption (similar to the
Brink hypothesis in electromagnetic transitions---see \cite{Aufetal})
that the energies in the resonance states scale in the same way as
those of the discrete states in the daughter, we adopt the mean value
of  8.7 MeV as the  $E_{IAS}$ in $^{54}$Fe corresponding to the g.s.
of $^{54}$Mn.  The numbers quoted in the last column of
Table~\ref{tab: M1comp} are for IAS states giving GT energy closest
to the experimentally observed GT-centroid .

 For the nuclei under consideration, the experimental inputs
coupled with the given theoretical prescriptions (M1 and FFN method)
give numbers which can be compared with the empirical relations
obtained here.
Tables~\ref{tab: FFNcomp} and~\ref{tab: M1comp} show that the
empirically fitted values of $E_{GT^+}$ and  $E_{GT^-}$ as reported
in section~\ref{sec:GTcentroid} are somewhat closer to the
experimental centroids than the numbers obtained from the M1 and FFN
methods. The results of
sections~\ref{sec: GTstrength} and~\ref{sec:GTcentroid} can therefore
be used to  calculate the weak transition properties of f-p shell
nuclei of interest to astrophysics.

\section{GT Collective States from Dispersion Relations  with Bohr-Mottelson
Hamiltonian}
\label{sec: CConst}
\noindent

Attempts have been made in the past to obtain a simple mass formula
for the GT$^-$ energy systematics using the single particle level
structure and two particle Bohr Mottelson type interaction
Hamiltonians.  The GTGR and the IAS states can be considered to be
collective excitations excited by the spin-dependent and
charge-exchange $\sum_{i=1}^A \tau_-^i \sigma^i$ and $\sum_{i=1}^A
\tau_-^i$ operators. More specifically, in the $GT^-$ direction, the
energy difference $E_{GT}-E_{IAS}$ can be written as ( see e.g.
Suzuki \cite{Suzetal}, 1982):
$$E_{GT^-}-E_{IAS}={{\langle \pi| \sum_{i=1}^A \tau_+^i \sigma _+^i H
                   \sum_{j=1}^A \tau_-^j \sigma _-^j | \pi \rangle}
              \over {\langle \pi | \sum_{i=1}^A \tau_+^i \sigma _-^i
                 \sum_{j=1}^A \tau_-^j \sigma _+^j | \pi \rangle}}
                  - {{\langle \pi | \sum_{i=1}^A \tau_+^i  H
                      \sum_{j=1}^A \tau_-^j | \pi \rangle}
                   \over
                    {\langle \pi | \sum_{i=1}^A \tau_+^i
                      \sum_{j=1}^A \tau_-^j | \pi \rangle}} $$
The following form of the two particle interaction Hamiltonian (Bohr
and Mottelson, 1981) for example has been used here by various
authors (e.g.  ref. \cite{Naketal}, \cite{Suzetal}, \cite{Suz81} ) :
$$ H= - \sum_{i=1}^A \xi_i l_i \sigma_i
   + {1 \over 2} \kappa_{\tau} \sum_{i \ne j} \tau_i . \tau_j
   + {1 \over 2} \kappa_{\sigma} \sum_{i \ne j} \sigma_i. \sigma_j
   + {1 \over 2} \kappa_{\sigma \tau} \sum_{i \ne j} (\tau_i . \tau_j)
                      (\sigma_i.\sigma_j) $$
where $\kappa_{\sigma}$, $\kappa_{\tau}$ and $\kappa_{\sigma \tau}$
are the coupling constants (divided by the mass number) for the
spin-spin, isospin-isospin and the spin-isospin interaction terms and
$\xi_i$ is the single particle spin-orbit coupling constant.  Using
the simplification brought about by considering the case of very
neutron rich nuclei where the Tamm-Dancoff approximation would be
valid and further considering only nulclei for which the protons
occupy the major shells and the neutrons fill the $j= l+1/2$
subshell, e.g. in the cases of $^{90}$Zr and $^{48}$Ca, Suzuki
\cite{Suz81} has used the experimentally known values of
$E_{GT^-}-E_{IAS}$ to determine the difference in the coupling
constants ($\kappa_{\tau} - \kappa_{\sigma \tau} $) (see Table 6).

Another approach (due to Gaarde \cite{Gaaretal}) to the GTGR energy
systematics is from the field description of the coherent
particle-hole excitations which lead to the GT resonance.  In the
field description, the coherent state is generated by an oscillation
of average field in the spin-isospin state which is propotional to
$\sigma \tau$. The self-consistency condition on this oscillating
potential leads to a dispersion relation for the energy of the GT
collective state. Further, if it is assumed that the strength is
clustered mainly around two regions, one corresponding to the
no-spin-flip transitions with energy $\epsilon_i$ and the other to
the spin-flip transitions with energy $\epsilon_i + \Delta_{ls}$
then, the energy dispersion relation equivalent to an  RPA  equation
for a seperable force can be written as:
$$ {{2(N-Z)(1-f)} \over {\epsilon_i - \epsilon }} + { {2(N-Z)f} \over
{\epsilon_i +\Delta_{ls}-\epsilon}} = {-1 \over {\kappa_{\sigma \tau}}}
    \eqno(6)$$
 where $f$ is the fraction of the total strength lying in the
spin-flip region, and $\Delta_{ls}$ is the spin-orbit splitting
energy. The quantity $f$ is evaluated for each of the nuclei (for
which (p,n) data exists) from the single particle transition
configurations constructed using the FFN method and is tabulated in
Table~\ref{tab: sf-fraction}.  A similar expression may be developed
for the IAS transition which has the form:
 $$\epsilon_{IAS} - \epsilon_{i,IAS}= 2 \kappa_{\tau} {(N-Z)}
  \eqno(7) $$
  Using the experimentally known  values of $E_{IAS}$ and eq. (7), we
obtain the value of $\kappa_{\tau}$ as $49.04/A$. Keeping this value
of $\kappa_{\tau}$ fixed, the  'experimental' values of $E_{GT}^-$
and $E_{IAS}$ obtained from Table~\ref{tab: FFNcomp} and Fig. 3 are
fitted by a two parameter least square fit to the quantity
$(\epsilon_{GT}-\epsilon_{i,GT})- (\epsilon_{IAS} -
\epsilon_{i,IAS})$, where the first term is evaluated from the
quadratic roots of eq. (6). This gives the value of $\kappa_{\sigma
\tau}$  as $11.16/A$.  As is shown in Table~\ref{tab: CConst}, these
values differ from those obtained for  $^{208}$Pb , or even for
$^{90}$Zr region, indicating that in order to explain both the IAS
and GT energetics, the simple Bohr-Mottelson Hamiltonian requires
coupling coefficients which are substantially different for different
shells.  At the same time it is not surprising that the relation (4)
developed for f-p shell nuclei which can be shown to depend on the
coupling constants (as in the approach of Suzuki \cite{Suzetal},
\cite{Suz81}), would differ substantially from similar relations
\cite{Naketal} developed for a different region of the isotope table.

\section{Conclusion}
\label{sec: Concl}
\noindent
Since the Spectral Distribution Theory is essentially a statistical
theory requiring the existence of a large number of basis states,
this approach is expected to be valid only for nuclei which are far
removed from the closed shell and sub-shell configurations.  Using
the framework of the Spectral Distribution Theory, we have used the
experimental data on charge exchange reactions on fp-shell nuclei to
obtain the energy  centroid of the collective Gamow-Teller resonances
for arbitrary nuclei in this shell. These quantities, together with
the prediction of total  GT strength ( such as given in
\cite{Koonetal} ) are useful in the prediction of reaction rates
mediated by weak-interactions in the cores of massive pre-supernova
stars.  We have in eqs. (4) and (5) a dependence of the GT energy
centroids in the f-p shell on  the spin orbit interaction, the Lane
potential and the pairing energy.  Further, with a view to comparing
the implications of the experimental results for fp-shell nuclei with
earlier work on heavier nuclei ( see \cite{Gaaretal}, \cite{Naketal}
and \cite{Suzetal} ) we have used the experimental values of the
GT$^-$ centroids and the IAS energies to extract the relevant
coupling coefficients in the Bohr-Mottelson Hamiltonian, for the f-p
shell.  The fact that they differ from those of the earlier studies
for heavier nuclei such as in the region of $^{208}$Pb indicates that
apart from the $1/A$ dependence, there may be an intrinsic shell
dependence of these coupling constants when a simple model like the
one due to Bohr-Mottelson has to be used.

The results of these calculations are being used to predict the
neutrino energy spectrum during the collapse phase upto the point of
$\nu_e$-trapping. In the event of a sufficiently nearby supernova
explosion, these can be compared with the neutrino spectroscopy
results obtainable by future neutrino detectors (e.g. SNO, ICARUS)
--- thereby revealing clues to the important thermodynamic and
nuclear conditions of the presupernova core. A preliminary discussion
of this was reported in ref.
\cite{Sutetal}.

We thank K. Kar and S. Sarkar for the initial impetus to work on this
problem. This research formed a part of the 8th Five Year Plan
($8{\rm P} - 45$) at Tata Institute.

Note added in the proofs:

 After this work was accepted for publication, a paper on experimental
G-T strength distributions in $^{60}$Ni, $^{62}$Ni and $^{64}$Ni by Williams
et al (Phys. Rev. C ${\bf 51}$, 1144 (1995)) came to our notice. Using their
experimental data, we find that the strength weighted energy centroids for
these nuclei are at 2.59, 2.56 and 2.8 MeV respectively. This is to be
compared with our predictions (from eq. (5)) of 2.59, 2.21 and 1.865 MeV.
While the predictions of eq. (5) (based on the (n,p) data of the earlier seven
nuclei) are reasonably good for $^{60}$Ni and $^{62}$Ni,
it does not work well in the case of $^{64}$Ni, possibly because
in the latter nucleus much of the strength is
concentrated in the ground state to ground state transition.

\begin{figure}
\caption{The best fit Edgeworth distribution superposed
on the $B_{GT}^-$ distribution obtained  from p-n reaction on
$^{71}$Ga at 120 MeV. The excitation energy $E_x$ is with respect
to the ground state of $^{71}$Ge.}
\end{figure}
\begin{figure}
\caption{ The best fit Edgeworth distribution superposed
on the $B_{GT}^+$ distribution obtained  from n-p reaction on
$^{59}$Co at 198 MeV.}
\end{figure}
\begin{figure}
\caption{Correlation between $E_{IAS}$ as calculated from eq. (3) and
the experimental positions of the IAS for fp-shell nuclei from (p,n)
reaction data. Nuclei indicated in the figure are ordered with
increasing values of experimental $E_{IAS}$.}
\end{figure}
\begin{figure}
\caption{$E_{GT^-}$ as obtained from eq. (4)  plotted against the
`experimental' $E_{GT^-}$ centroids. Nuclei indicated are given in
order of increasing value of experimental $E_{GT^-}$.}
\end{figure}
\begin{figure}
\caption{$E_{GT^+}$ as obtained from eq. (5)  plotted against the
`experimental' $E_{GT^+}$ centroids. Nuclei are given in
order of increasing value of experimental $E_{GT^+}$.}
\end{figure}
\begin{figure}
\caption{Schematic diagram of single particle excitations
illustrating the application of FFN method to determination of
$GT^- $ strength centroid with respect to the g.s. of $^{51}$Cr
in $\beta^-$-decay of $^{51}$V. The level spacings are not drawn
to scale.}
\end{figure}
\begin{figure}
\caption{ The M1 method for $^{54}$Fe(e$^-$,$\nu_e$)$^{54}$Mn.
Fig.7a shows the g.s.  configuration of $^{54}$Fe. In Fig. 7b,
the diagram on the left represents a spin-flip excitation in
in $^{54}$Mn generated from $^{54}$Fe g.s. by transforming a
$1_{f_{7/2}}$ proton into a $1_{f_{5/2}}$ neutron. The IAS for
the $^{54}$Fe configuration is a $\{\protect\sqrt{3/4},
\protect\sqrt{1/4}\}$ superposition of the shown basis states
on the right; the AIAS state is similarly $\{-\protect\sqrt{1/4},
\protect\sqrt{3/4} \}$. The AIAS is seperated from the IAS
only by the Lane potential.}
\end{figure}

\begin{table*}
\begin{center}
\caption{ Edgeworth parameters for $B_{GT}^-$ energy distribution
obtained from p-n reactions. \label{tab: Edge-} }
\vskip 1 true cm
\begin{tabular}{ccccc}
Nucleus& $A_0$($B_{GT}^-/{\rm MeV}$) & $E_{GT}^-$ (MeV)
& $\sigma$ (MeV) & $\gamma_1$ \\
\tableline
$^{51}$V & 1.2 & 10.7 & 4.2 & -1.00 \\
$^{81}$Br & 2.0  & 10.2 & 3.6 & -1.19 \\
$^{71}$Ga & 2.0 & 9.6 & 3.0 & -1.25 \\
$^{58}$Ni & 1.0 & 9.4 & 1.8 & 0.11 \\
$^{60}$Ni & 0.8 & 9.0 & 2.6 & 1.05 \\
$^{54}$Fe & 1.2 & 8.9 & 1.8 & -0.50 \\
$^{56}$Fe & 1.0 & 9.0 & 3.5 & -0.67 \\
\end{tabular}
\end{center}
\end{table*}
\vfill
\eject
\begin{table*}
\begin{center}
\caption{Edgeworth parameters for $B_{GT}^+$ energy distribution
obtained from n-p reactions data.\label{tab: Edge+}}
\vskip 1 true cm
\begin{tabular}{ccccc}
Nucleus& $A_0$ ($B_{GT}^+/$MeV)& $E_{GT}^+$ MeV & $\sigma$ (MeV)
& $\gamma_1$ \\
\tableline
$^{54}$Fe & 0.6 &2.8 & 2.1 & 1.01 \\
$^{51}$V & 0.3 & 3.8 & 1.3 & -1.21 \\
$^{59}$Co & 0.6 & 4.3 & 1.2 & 0.43 \\
$^{70}$Ge & 0.1 & 2.1 & 2.1 & -1.53 \\
$^{56}$Fe & 0.6 & 2.3 & 1.8 & 0.97 \\
$^{55}$Mn & 0.4 & 4.4 & 1.9 & 0.53 \\
$^{58}$Ni & 1.0 & 3.3 & 1.5 & 0.59 \\
\end{tabular}
\end{center}
\end{table*}
\vfill
\eject
\begin{table*}
\begin{center}
\caption{ Experimental values of the energy-centroids in MeV for
$GT^-$ excitation in the mother nucleus and values obtained from the
fit eq.(4) and from the FFN Method.\label{tab: FFNcomp}}
\vskip 1 true cm
\begin{tabular}{cccc}
Nucleus& $E_{GT^-}$ (MeV) & $E_{GT^-}$ (MeV) & $E_{GT^-}$ (MeV) \\
&(Expt.) &(Eq. 4) &(FFN)\\
\tableline
$^{51}$V  & 10.7 & 11.1 & 10.6  \\
$^{81}$Br & 10.2 & 9.6 & 12.0 \\
$^{71}$Ga & 9.6 & 9.9  & 13.6 \\
$^{54}$Fe & 8.9 & 8.9  & 8.7\\
$^{58}$Ni & 9.4 & 9.0  & 6.5\\
$^{56}$Fe & 9.0 & 9.7  & 12.6\\
$^{60}$Ni & 9.0 & 8.7  & 4.9\\
\end{tabular}
\end{center}
\end{table*}
\vfill
\eject
\begin{minipage}[t]{6 in}
\begin{table*}
\begin{center}
\caption{ Experimental values of the energy-centroids for
$GT^+$ excitation in the mother nucleus and the values obtained from
eq. (5) and from the M1 method.\label{tab: M1comp}}
\vskip 1 true cm
\begin{tabular}{ccccc}
Nucleus & $E_{GT^+} (MeV)$ & $E_{GT^+}$ (MeV)& M1 config.$^a$
(daughter) &$E_{GT^+}$ (MeV)\\
& (Expt.) & (Eq. 5)&&(M1)$^b$\\
\tableline
$^{54}$Fe & 2.8 & 3.05
&($\pi^5_{f_{7/2}}$),($\nu_{f_{5/2}}, 0_{p_{3/2}}$)&5.17(4.07) \\
$^{56}$Fe & 2.3 & 2.62
&($\pi^5_{f_{7/2}}$),($\nu_{f_{5/2}}, \nu^2_{p_{3/2}}$) & 4.44
(4.60)\\
$^{59}$Co& 4.3& 3.97
&($\pi^6_{f_{7/2}}$),($\nu_{f_{5/2}}, \nu^4_{p_{3/2}}$) &6.29(5.34)\\
\end{tabular}
\end{center}
\vskip -0.5 true cm
\noindent
(a) The single particle shell model configurations \cite{Hill} for
the outermost levels of the daughter in the M1 excited state
(Seeger neutron levels for both protons and neutrons). The
levels are given in order of increasing energy, starting with the
lowest energy, partially-filled obital.

\noindent
(b) The numbers quoted within parenthesis are the $E_{GT}^+$
    values from M1 method when the seperate 'Seeger' \cite{Hill}
    levels are used for neutrons and protons.
\end{table*}
\end{minipage}
\vfill
\eject
\begin{table*}
\begin{center}
\caption{ The distribution of $B_{GT}^-$ strength in the spin-flip
(sf) and no-spin-flip (nsf) transitions. Column 4 gives the
fractional strength of the sf transitions.\label{tab: sf-fraction} }
\vskip 1 true cm
\begin{tabular}{cccc}
Nucleus& $B_{GT}^-$ (nsf) & $B_{GT}^-$ (sf) & f \\
\tableline
$^{51}$V & 6.42 & 13.71 & 0.68 \\
$^{81}$Br & 10.13 & 14.47 & 0.59\\
$^{71}$Ga & 6.60 & 15.04 & 0.69 \\
$^{54}$Fe & 2.57 & 13.71 & 0.84 \\
$^{58}$Ni & 3.33 & 16.37 & 0.83 \\
$^{56}$Fe & 5.90 & 16.37 & 0.73\\
$^{60}$Ni & 0.0 & 19.04 & 1.00 \\
\end{tabular}
\end{center}
\end{table*}
\vfill
\eject
\begin{minipage}[t]{6 in}
\begin{table}
\begin{center}
\caption{ Coupling constants $\kappa_{\tau}$, $\kappa_{\sigma
\tau}$ for Bohr-Mottelson two particle interaction.
\label{tab: CConst}}
\vskip 1 true cm
\begin{tabular}{ccccc}
Authors& Nuclei & $A \kappa_{\tau}$ & $A \kappa_{\sigma \tau}$ &
$A(\kappa_{\tau}- \kappa_{\sigma \tau})$ \\
\tableline
Gaarde \cite{Gaaretal} & $^{208}$Pb & 28 & 23 & 5 \\
Suzuki \cite{Suz81}& $^{90}$Zr & 32.5$^a$
& 25.9 - 28.6   &3.92\\
Suzuki \cite{Suz81}& $^{208}$Pb & 32.5$^a$
& 28.1 -29.0$^a$  &3.95\\
 Nakayama & `Global' fit & 32.5 & 23.25 & 9.25 \\
et al. \cite{Naketal}& $^{92}$Zr to $^{208}$Pb & & & \\
 This work & f-p shell & 49.0 & 11.2 & 37.8 \\
\end{tabular}
\vskip 0.5 true cm
\leftline{(a) Bohr-Mottelson estimate, quoted in \cite{Suz81}.}
\end{center}
\end{table}
\end{minipage}
\vfill
\eject
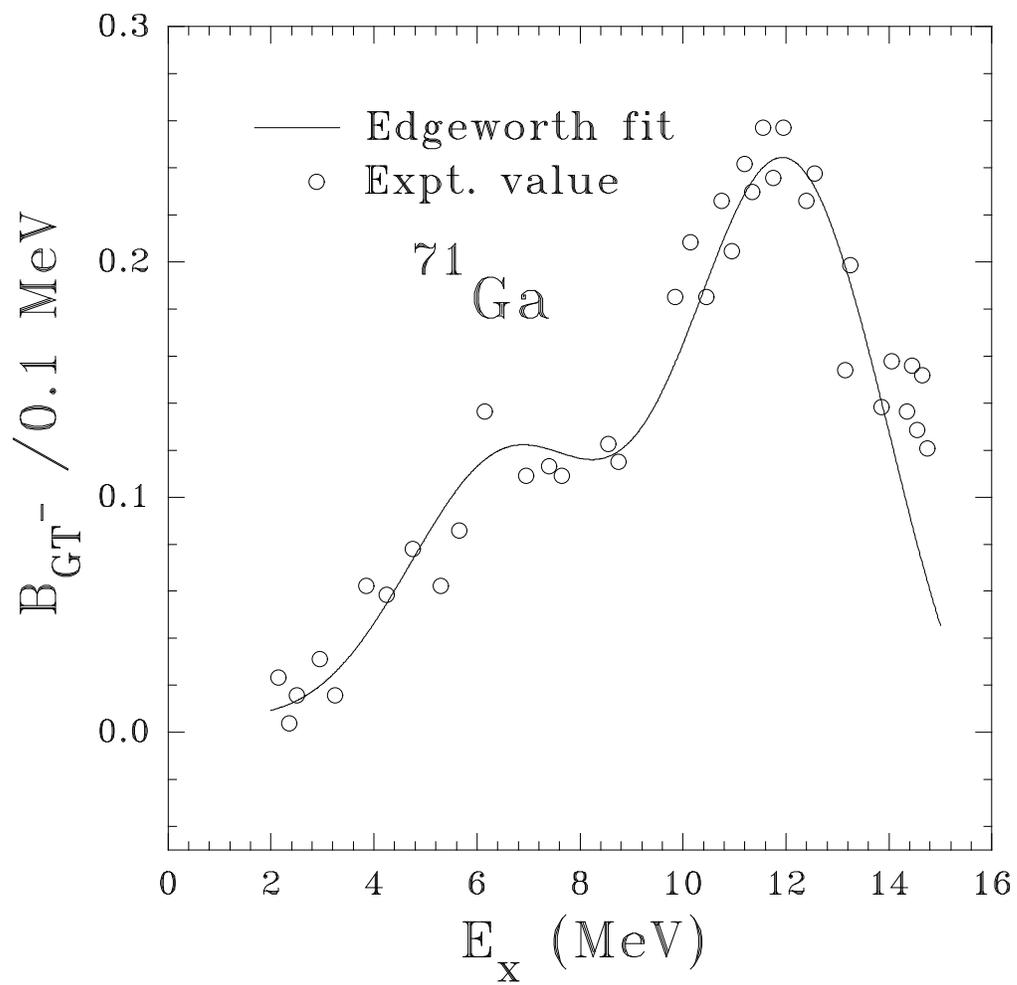
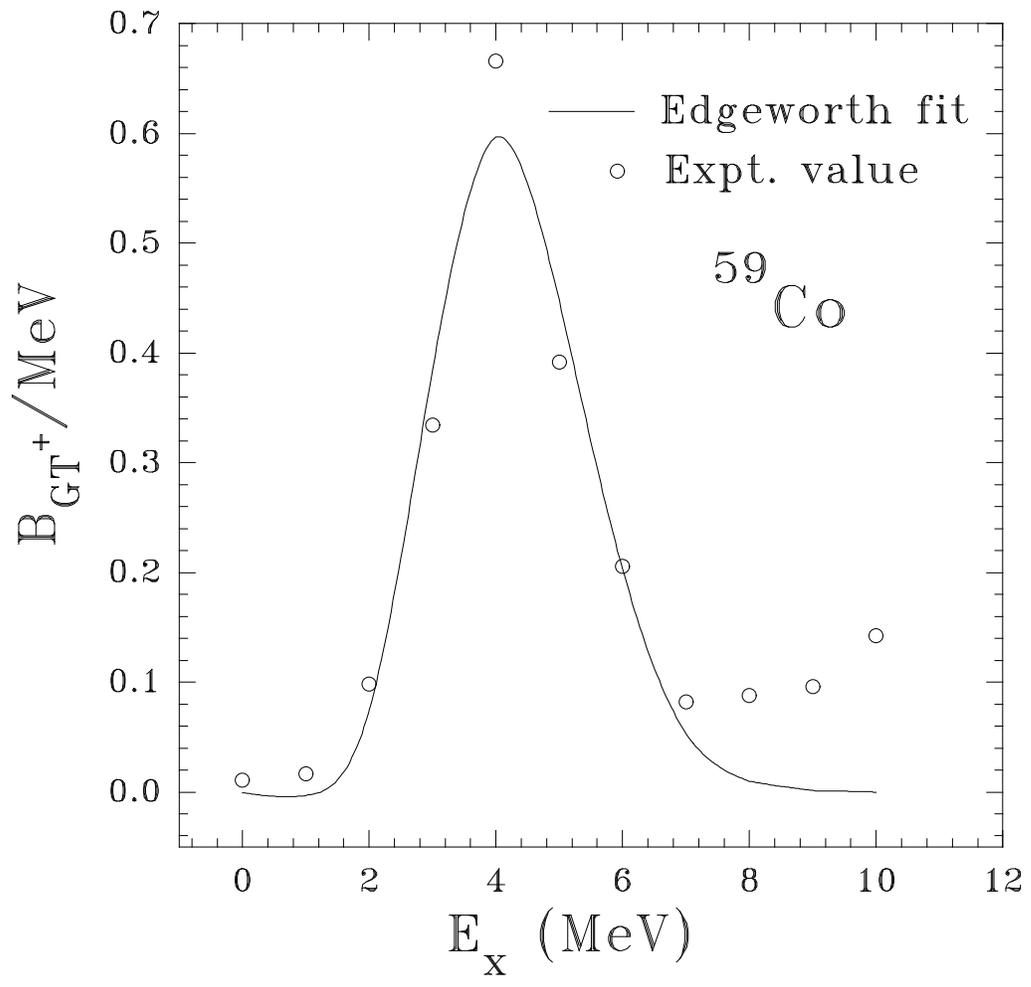
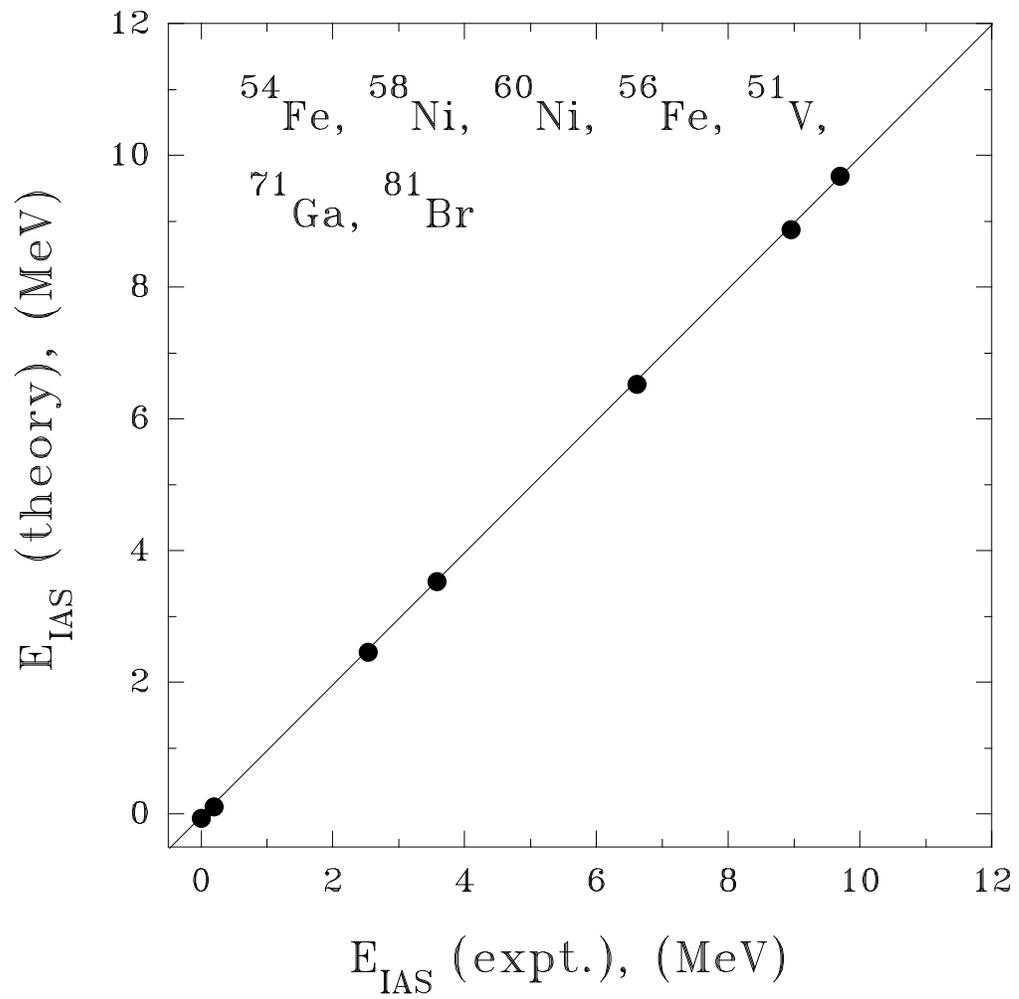
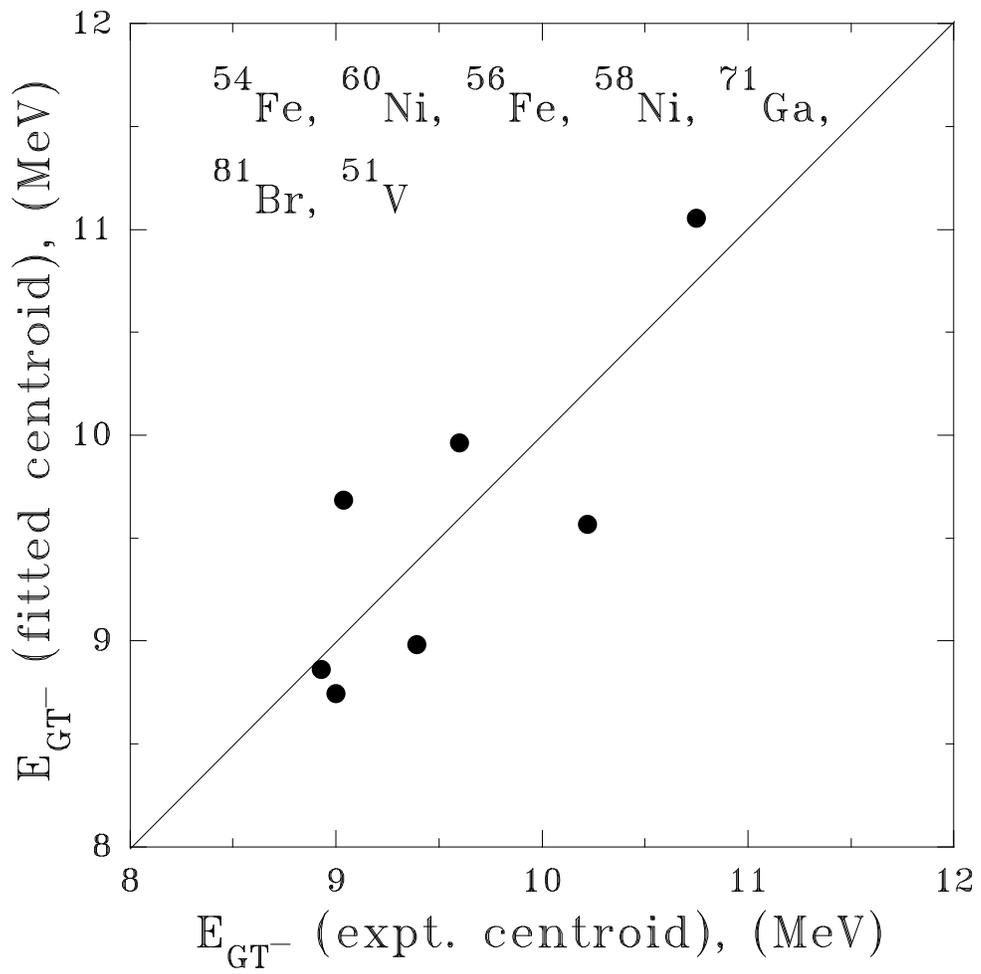
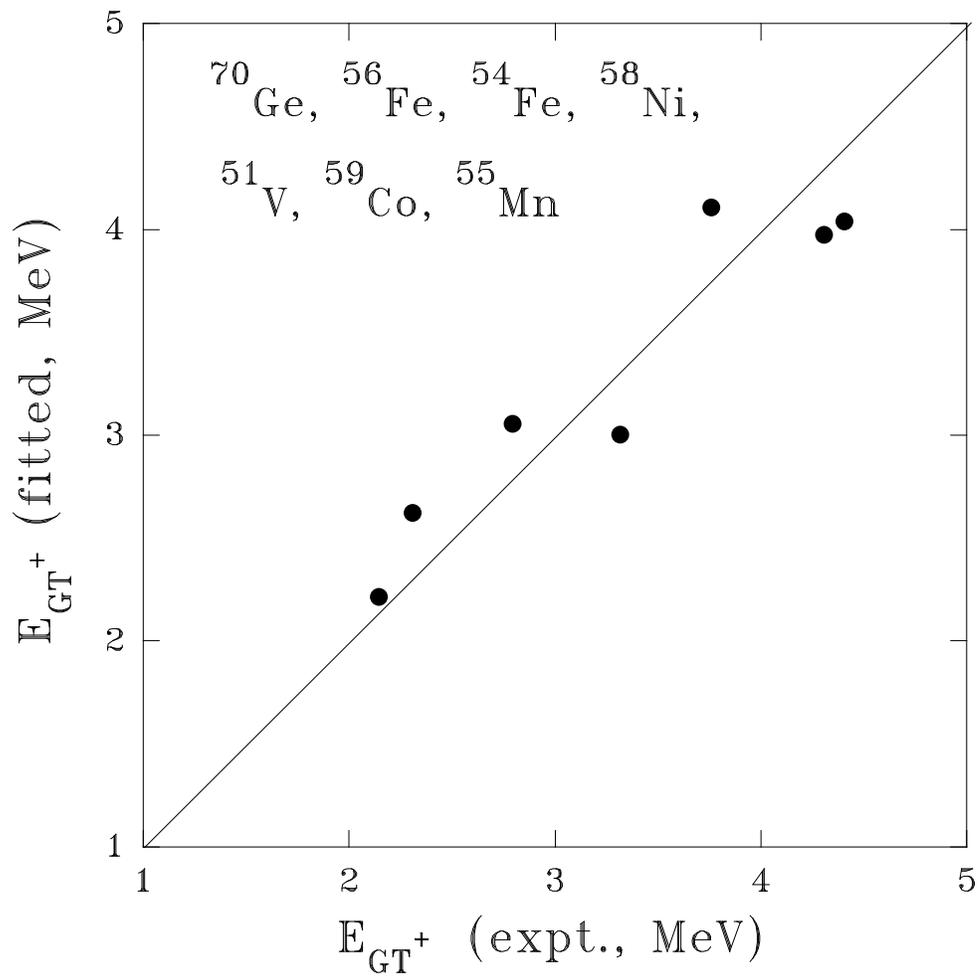
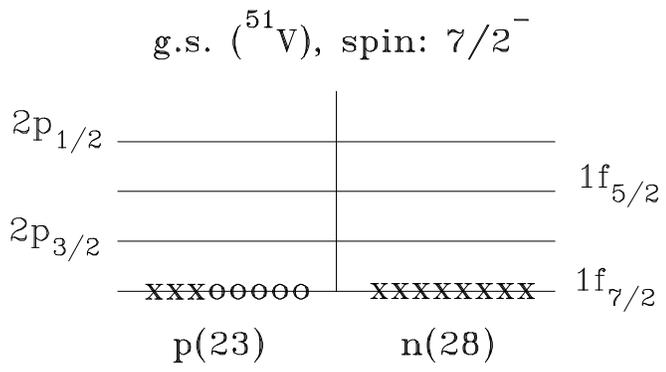
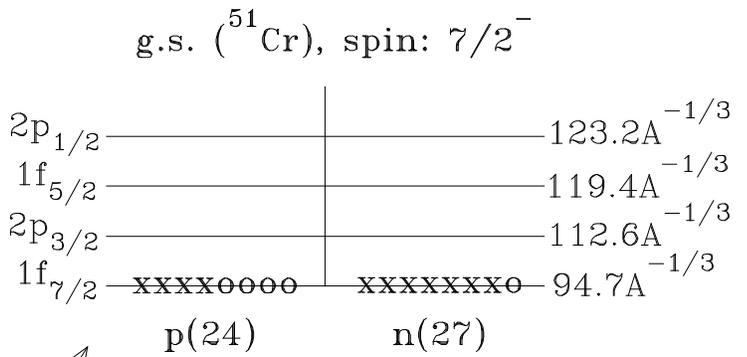
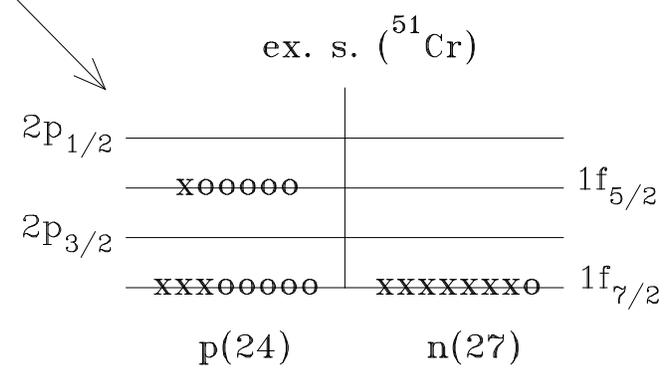
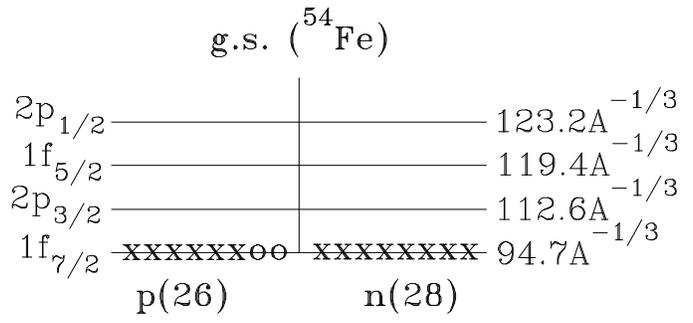
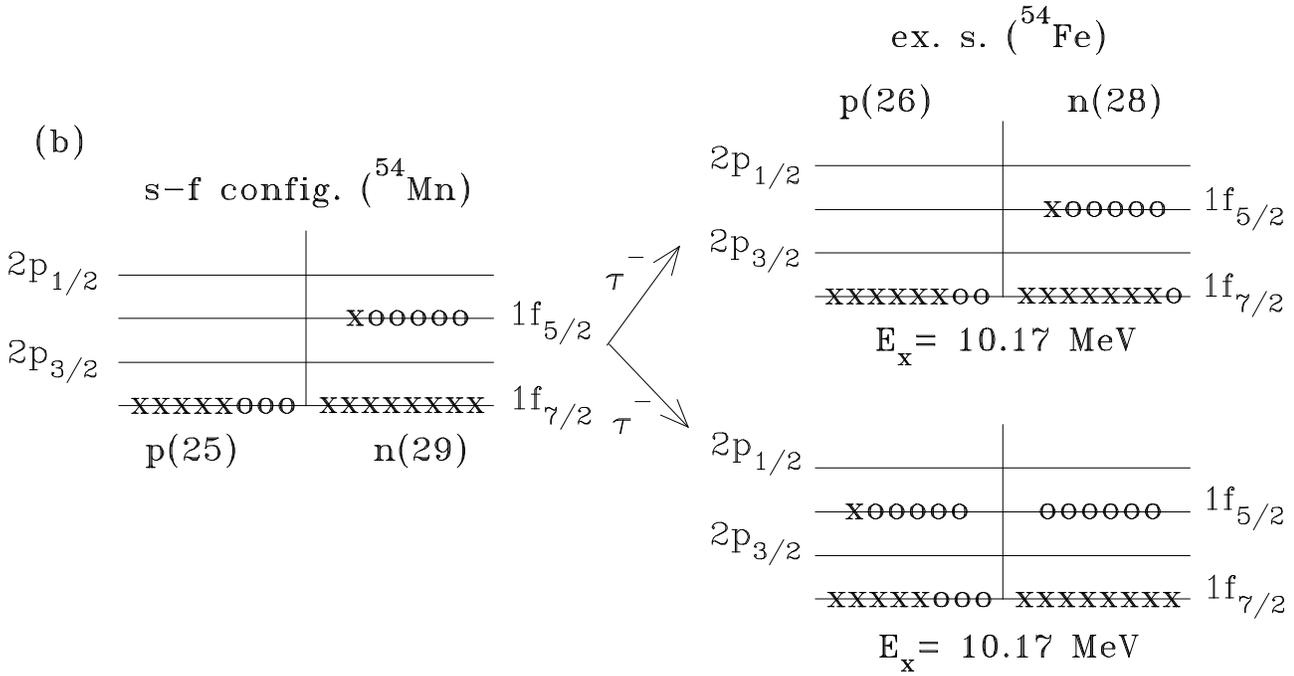
\end{document}